\documentclass{aastex}
\usepackage{emulateapj5}

\input psfig.sty
\def\degree{$^{\circ}$}

\def\tlabel#1{\textit{#1}}
\def\vkm{km s$^{-1}$}
\def\smyr{M$_\odot$ yr$^{-1}$}
\def\sl{$L_\odot$}
\def\cm3{cm$^{-3}$}
\slugcomment{\today}

\def\epsfbox#1{\plotone{#1}}
\def\epsfbox#1{}    	  
\def\clearp{\clearpage}
\def\clearp{}
\def\leftblank#1{}   
\def\cdir{.}
\def\SIIratio{[SII]$\lambda6716$\AA/$\lambda6730$\AA{}}

\def\ifEqString#1#2{\def\testa{#1}\def\testb{#2}%
    \ifx\testa\testb}

\begin{document}

\title{Shaping Proto-Planetary and Young Planetary Nebulae with Collimated
Fast Winds}

\author{Chin-Fei Lee and Raghvendra Sahai}
\affil{Jet Propulsion Laboratory, MS 183-900, 4800 Oak Grove Drive,
Pasadena, CA 91109}
\email{chinfei@eclipse.jpl.nasa.gov, sahai@eclipse.jpl.nasa.gov}

\begin{abstract}
Using two-dimensional hydrodynamical simulations, 
we investigate the interaction of a collimated
fast wind (CFW) interacting with a spherical asymptotic giant branch (AGB)
wind as the mechanism 
for shaping proto-planetary nebulae and young planetary nebulae.
In particular, we compare our simulations 
to the observations of an evolved PPN with multiple, 
highly collimated lobes, CRL 618. 
We characterize our model CFW by
three parameters: opening angle, velocity and mass-loss rate, and explore the 
dependence of the properties of the shell on the first two.
For given opening angle and velocity, the mass-loss rate 
is chosen to give a shell velocity of about 150 \vkm{} at
the tip, similar to that seen in CRL 618.
In our simulations, the shell dynamics is found to depend on
the velocity of the fast wind: we obtain a momentum-driven shell for a 300
\vkm{} fast wind and a ballistic bow-shock driven shell 
for a 1000 \vkm{} fast wind.  
The shell driven by the collimated fast wind is highly collimated, even 
though the AGB wind is spherical. 
Time variations in the velocity of the fast wind
produce a series of internal shock pairs
interacting with the inner surface of the shell. Due to radial expansion,
the density of the internal shocks decreases with distance.

Various emission diagnostics have been derived from our simulations.
For a 300 \vkm{} fast wind, the optical emission arises from both
the shocked AGB wind and shocked fast wind, showing 
one or two bright bow-like structures at the tip of the lobe. However,
for a 1000 \vkm{} fast wind, since the shocked fast wind is much hotter, 
it emits mainly in X-ray emission; 
the optical emission forms only one bow-like structure at the tip associated
with the shocked AGB wind. 
The position-velocity (PV) diagrams derived from our simulations
all show a broad range of velocities at the tip.
The detailed PV structure and velocity range at the tip 
depend on the shell dynamics and the relative contributions of the shocked
fast wind and shocked AGB wind.

We make a detailed comparison of our simulations
to the observations of the relatively isolated northwestern (W1)
lobe of CRL 618. We find that a 300 \vkm{} collimated fast wind 
with an opening angle of 10\degree{}
can readily produce a highly collimated lobe similar to the W1 lobe,
including the bow-like emission structure at its tip.
However, our models have difficulty producing
the bright emission structures seen along the body of the lobe.
The \SIIratio{} ratios at the tip of the lobe in all of our simulations 
are similar to that observed at the tip of the W1 lobe.
The optical line ratios indicate a temperature stratification in the tip, both for the
simulations and observations, however, the temperatures at the tip of the lobe
in our simulations are higher than observed. 
The position-velocity (PV) diagrams derived from our simulations
are all qualitatively consistent with the current observations. 
The collimated fast wind in CRL 618 is unlikely to be steady and is not radiatively
driven.
\end{abstract}

\keywords{stars: AGB and post-AGB --- stars: winds and outflows.}

\section{Introduction}

An investigation of the physical mechanism(s) responsible for 
the shaping of proto-planetary nebulae (PPNs) and young planetary nebulae (PNs)
is crucial for understanding the mass-loss processes during the end 
phases of the evolution of low- and intermediate-mass stars.
During the asymptotic giant branch (AGB) phase, 
these cool stars undergo intense mass-loss (with rate of up to $10^{-4}$ \smyr{}) via a dense, slow 
($\sim$ 10 - 20 \vkm{}) wind \cite[see e.g., review by][]{Balick2002}, 
forming a circumstellar envelope (CSE) of dust and gas around 
the star. 
During the PN phase, the central stars (now hot white dwarfs) lose mass via 
fast winds at  speeds of $\sim$1000-2000 \vkm{}
and rates of $10^{-8}$ to $10^{-6}$ \smyr{} \citep{Balick2002}.
The fast wind catches up with the AGB wind and drives a shocked shell 
through it, forming a PN.
This is the so-called interacting stellar winds (ISW) model
commonly used to explain the formation of PNs
\citep{Kwok1978,Kahn1983,Balick1987B,Frank1994,Dwarkadas1996,Frank1999}.

In the ISW model, the fast winds are spherical, 
and the AGB CSEs are assumed to have
a density higher near the equator than near the pole,
so that the shells driven by the 
fast winds expand fastest along the nebular symmetry axis.
By varying the ratio of equatorial to 
polar density in the CSE, the ISW model is able to produce 
a variety of axisymmetric (e.g. elliptical or bipolar) shaped PNs
\citep{Balick1987B,Frank1994}.

However, recent high spatial resolution and high dynamic-range imaging 
of many PPNs and young PNs with the Hubble Space Telescope (HST) shows 
that almost all of these have highly aspherical shapes, with a significant fraction 
having highly collimated bipolar or multipolar lobes \citep{Sahai2001,SahTrg98}. 
Examples of objects with highly collimated lobes were also known from
previous ground based observations \cite[e.g., M 2-9,  M1-91, M1-16,][]
{Corradi1995,Schwarz1997}.
Point symmetry, rather than axisymmetry, better characterizes the geometry of the 
majority of these objects [striking examples are He\,2-115 \citep{SahTrg98},
IC\,4634 \citep{Sahai2003}, and He\,3-1475 \citep{Borkowski1997}]. 
In view of such observations,
\cite{SahTrg98} concluded that the ISW model could not explain the shaping of PNs, 
and proposed that collimated fast winds operating at the beginning of the 
post-AGB phase were the primary mechanism for initiating the shaping of 
PNs.

Several nagging problems for the ISW model existed even 
before the HST imaging results 
became available. E.g., the shocked fast winds in the ISW model (without
magnetic fields) did not converge to form stable,
low-ionization microstructures \citep{Icke1992,Frank1994,Dwarkadas1998} 
like ansae, thin jets, or FLIERs
\citep{Balick1993}, 
since at the high speeds of the fast winds, most 
windblown bubbles are ``energy conserving" \citep{Dyson1980}.
MHD models with a toroidal magnetic field embedded in a normal radiatively driven
stellar wind are found to be able to produce such microstructures
\citep{Rozyczka1996,Garcia1997}. However, the initial conditions in
these models may not be appropriate \citep{Gardiner2001}. In addition,
these models require radiatively driven winds which 
can not account for the very large momentum excesses observed in PPNs
\citep{Bujarrabal2001}.
Moreover, in order to produce bipolar PNs with collimated lobes and narrow waists,  
very high ratios of equatorial to polar densities of the AGB winds were 
needed \citep{Balick1987B,Mellema1995}. 
However, less than 20\% of the AGB envelopes show significant
deviations from spherical symmetry \citep{Bujarrabal1991,Neri1998}. 
Faint, roughly round halos can be seen around 
bright PPNs and young PNs, indicating that their progenitor AGB envelopes are
produced by isotropic mass-loss \citep{Chu1987,Sahai2001}.
The dense equatorial waist observed 
in many bipolar PPNs \cite[e.g.,][]{Sahai1998,Kwok1998,Su1998} may actually result 
from the interaction between a collimated fast wind 
and the spherical AGB wind \citep{Soker2000}.

Thus, the evidence for the presence of CFW's in dying stars seems overwhelming, yet very little work 
has been done so far in order to investigate in detail how CFWs can shape the circumstellar 
environment in these objects, and to constrain the physical properties of the CFWs through quantitative 
comparisons with observations. This paper makes a first attempt at
addressing the above lack. We explore the interaction of a CFW with a spherical AGB wind using 
two-dimensional hydrodynamical simulations.
The CFW is assumed to emanate radially from the source
with a density decreasing from the pole to the equator.
It is thus structurally different from a {\it cylindrical} jet,
which emanates along the outflow axis with
constant diameter and density.  
Cylindrical jet models are much more common in the literature.
These have been explored extensively in the case of 
YSOs \citep{Blondin1990,Stone1994,Biro1994,Raga1995,Suttner1997}, 
and to a much lesser extent, for PNs 
\citep{Cliffe1995,Steffen1998}. A CFW model has been
used to produce the bipolar lobes associated with 
the luminous blue variable, $\eta$ Car \citep{Frank1998}.

We support the results of our numerical simulations by comparison 
with analytical outflow models, and derive observational diagnostics for 
comparison with observations.
In \S2 and 3, we describe the parameters and the numerical method
used in the simulations. 
In \S4, we present our simulations and compare them 
with analytical outflow models.
In \S5, we compare our simulations with the observations 
of a young PPN, CRL 618.
We summarize and conclude our work in \S6.

\section{Numerical Simulation Parameters}

In our simulations, the AGB wind is assumed to be
spherically symmetric with the density given by
(in spherical coordinate)
\begin{equation}
\rho_a = \frac{\dot{M}_{a}}{4 \pi r^2 v_a}
\end{equation}
where $\dot{M}_{a}$ and $v_a$ are the mass-loss rate 
and velocity of the AGB wind, respectively. 
For our simulations to be compared with the observations of CRL 618, 
$\rho_a$ is calculated with
$\dot{M}_{a}=3\times10^{-5}$ \smyr{} and $v_a=20$ \vkm{},
the values comparable to those for CRL 618 \citep{Knapp1985}.
After $\rho_a$ is calculated, however,
$v_a$ is set to zero because it is very small compared to the velocity 
of the fast wind. 
This assumption of $v_a$ is appropriate in this paper,
because we focus on the shell dynamics away from the equatorial
regions. The temperature of the AGB wind is low and set to 10 K.

At present, there is no consensus on how the fast wind 
is launched and collimated.
One possibility is that 
the fast wind is launched from a magnetized accretion
disk and star system 
by the magneto-centrifugal forces, as proposed for AGN jets 
\cite[e.g.,][]{Blandford1982} and protostellar winds/jets 
\cite[e.g.,][]{Shu1994,Konigl2000}.
These winds/jets are highly collimated by the magnetic fields.
Keeping in mind our lack of understanding of the physical mechanisms
driving the fast wind, we assume a {\it simple} parametrized description of
the CFW density and velocity structure:
\begin{eqnarray}
\rho_f &=& \rho_{fo}\exp\;[-(\theta/\theta_f)^2] \nonumber \\
{\bf v}_f &=& v_{fo}\exp\;[-(\theta/\theta_f)^2] \;{\bf \hat{r}}
\end{eqnarray}
where $\rho_{fo}$ and $v_{fo}$ are the density and velocity at the pole
(where $\theta=0$\degree), and $\theta_f$ is the opening angle.
The momentum per unit volume of the fast wind has a full width
at half maximum of $\sqrt{2\ln 2} \theta_f = 1.18 \theta_f$.
For a given mass-loss rate of the fast wind (into two opposite
sides of the central source),
$\dot{M}_f$, the density at the pole is
\begin{equation}
\rho_{fo}=\frac{\dot{M}_f}{4 \pi r_f^2 v_{fo} K}
\end{equation}
where $r_f$ is the launching radius of the fast wind and
\begin{equation}
K=\int_0^{\pi/2} \exp[-2(\frac{\theta}{\theta_f})^2] \sin \theta
\;d\theta
\end{equation}
In our simulations,  $\rho_{fo}$ is calculated from $\dot{M}_f$ with
$r_f$ assumed to be 5$\times10^{15}$ cm (333 AU), so that
our fast wind is characterized by three parameters: 
$\dot{M}_f$, $v_{fo}$ and $\theta_f$.
Originating near or from the post-AGB star, the fast wind is assumed to have a
temperature of $10^4$ K.

The parameters of the fast wind in our simulations are listed in 
Table \ref{tab:parameters}.
Observationally, the fast wind is found to have a velocity 
up to $\sim$ 2000 \vkm{} \cite[e.g., He 3-1475,][]{Sanchez2001}.
Here, $v_{fo}$ is set to
two velocities, 300 and 1000 \vkm{}. 
In the simulation with  $v_{fo} =$ 1000 \vkm{},
$v_{fo}$ is allowed to increase linearly 
from 0 to 1000 \vkm{} in the first 10 years 
to avoid a very strong shock at the beginning.
As we will see later,
these two velocities of $v_{fo}$ result in qualitatively different 
dynamics in the simulations.
$\theta_f$ is set to 10\degree{} and 20\degree{}
in order for our simulations to be compared 
with highly collimated PPNs and young PNs.
$\dot{M}_f$ ranges from 2.5$\times10^{-7}$ to $10^{-5}$ \smyr{}. 
It is chosen to give a 
shell velocity at the tip of about 150 \vkm{},
which is comparable to that seen in CRL 618 \cite[][hereafter SCSG02]{Sanchez2002}.

The fast wind may have periodical variations in density and velocity,
as suggested in e.g., CRL 618 (SCSG02). Such variations 
could result from variations in the accretion process
as suggested for protostellar winds \citep{Hartmann1996}.
In this case, the density and velocity are assumed to vary in such a way
that the mass-loss rate of the fast wind is constant over the variation.
Hence, the fast wind is assumed to have
\begin{eqnarray}
\rho_f' &=& \rho_f/(1+A \sin \frac{2\pi t}{P}) \nonumber \\
{\bf v}_f' &=& {\bf v}_f(1+A \sin \frac{2\pi t}{P})
\label{eq:ft}
\end{eqnarray}
where $A$ and $P$ are the amplitude and period of the variation.
Such variations in density and velocity 
have also been used in simulations of protostellar
winds/jets in order to produce knotty and bow-like structures 
along the outflow axis
\cite[see e.g.,][]{Biro1994,Lee2001}.

\section{Equations and Numerical Method} \label{sec:method}
The two-dimensional hydrodynamic code, 
ZEUS 2D, is used to solve the equations of hydrodynamics, 
\begin{eqnarray}
\frac{\partial{\rho}} {\partial{t}}+ \nabla\cdot(\rho{\bf v}) &=&0 \nonumber\\
\frac{\partial{\rho {\bf v}}} {\partial{t}}
+\nabla\cdot(\rho {\bf v}{\bf v}) &=&
-\nabla p + f \nonumber\\
\frac{\partial{e}} {\partial{t}}
+\nabla\cdot(e{\bf v}) &=&
-p\nabla\cdot{\bf v}-n^2\Lambda
\end{eqnarray}
where $\rho$, ${\bf v}$, $p$, $e$, and $n$ are the mass density,
velocity, thermal pressure,
internal energy density, and hydrogen nuclei number density,
respectively. 
An ideal gas equation of state $p=(\gamma-1)e$ 
with $\gamma=5/3$ is used for the thermal pressure.
Helium is included as a neutral component
with $n(\textrm{He})=0.1\cdot n$, so that $n=\rho/(1.4\cdot m_H$),
where $m_H$ is the mass of atomic hydrogen.
In our simulations, the undisturbed AGB wind is assumed to be static.
As a result, a small fixed force per unit volume, $f$, is introduced 
in our simulations to
hold the undisturbed AGB wind from thermal expansion.
Here $f = -\frac{2 p_a}{r}$, where $p_a$ is the thermal
pressure of the undisturbed AGB wind.
$\Lambda$ is the optically thin
radiative cooling function for interstellar gas, with 
the function at high temperature from \citet{MacDonald1981} and 
the function at low temperature from \citet{Dalgarno1972}.
We assume an equilibrium cooling function, in which the
ionization fraction of hydrogen is calculated by equating the ionization
rate with the recombination rate of hydrogen.
As a result, the cooling function decreases rapidly at about 10$^4$ K
as the ionization fraction of hydrogen decreases.
Molecular cooling is not included. 

Scalar tracers, $c$, $c_a$, and
$c_w$, are also included in the simulations to track
the fast and AGB winds.
$c$ tracks the fast wind, 
it is one for the fast wind,
zero for the AGB wind, and a value between one and zero 
for a mixture of the fast wind and AGB wind.
$c_a$ and $c_w$ trace respectively 
the flows of the AGB wind and the fast wind 
in the strongly interacting regions near the pole.
Thus, $c_a$ ($c_w$) 
is one for the AGB (fast) wind emanating within the angular
segment 5\degree{}$<\theta<$7.5\degree{}, a value between one and zero
for a mixture of this material with the rest, 
and zero for the rest. 

Simulations are performed in spherical coordinates with
a computational domain of dimensions 
$(r,\theta)=( 5\times10^{15} \;\textrm{to}\; 10^{17} 
\;\textrm{cm},\; 0 \;\textrm{to}\; \theta_b)$ 
and a uniform grid of $400\times400$ zones.
This gives a resolution of $2.37\times10^{14}$ cm in $r$.
$\theta_b$ is a value between 0.5 and $\pi/2$ depending on the angular
extent of the simulations,
giving a resolution of 1.25$\times10^{-3}$ to 4$\times10^{-3}$ radians in $\theta$.
Reflecting boundary conditions are used along the inner $\theta$
(where $\theta=0$\degree{})
and outer $\theta$ (where $\theta=\theta_b$) boundaries, 
while outflow boundary conditions are used along the
outer $r$ boundary. 
Inflow boundary conditions are used for inner $r$ boundary 
to introduce the fast wind into the AGB wind.
The results of the simulations are presented in cylindrical coordinates
(z,R), with the z-axis being the outflow axis.

\section{Results}

In this section, we present the results of our simulations
as the fast wind propagates to a distance of about 10$^{17}$ cm,
a length similar to that of the lobes in CRL 618.
Observationally, a significant amount of optical emission is 
believed to come from a shell surrounding the fast wind, therefore,
we concentrate on the shell dynamics in the simulations.
Model 1 is considered as a standard model because it can
produce a shell with a collimation similar to that of CRL 618.
We discuss the effects of changing the opening angle and velocity
of the fast wind on the structure and kinematics of the shell.
We use analytical models to support the results of our simulations.
At the end of this section, we discuss how the velocity of the fast wind
changes the shell dynamics in the simulations.

\subsection{Model 1: Standard model}\label{sub:model2}

The simulation of model 1 is presented in Figure \ref{fig:model2}.
The fast wind is 177 years old.
In this model, $\dot{M}_f=2.5\times10^{-6}$ \smyr{},
$v_{fo}=300$ \vkm{}, and $\theta_f=10$\degree{}.
As the fast wind blows into the AGB wind, a shell  is formed around the fast wind.
The shell is highly collimated
with a length to width ratio of about 6.
It consists of shocked AGB wind (forward shock)
and shocked fast wind (backward shock) separated 
by a contact discontinuity (mixture), as indicated by the $c$ contours.
Since the heat generated by the fast wind's kinetic energy is
efficiently converted into escaping radiation,
the shocked fast wind becomes
a thin inner shell accelerating the shocked AGB wind in the outer shell. 
As a result,
the shell is a ``momentum-conserving'' wind-driven shell.
The motions of the shocked fast wind and shocked AGB wind are different.
The shocked fast wind moves along the shell toward the pole,
producing a focusing effect that in
turn produces a high-density jet-like structure 
at the tip, as indicated by the $c_w$ contours \cite[see][]
{Canto1988,Frank1996b}. The detailed structure of the jet-like structure,
however, depends on the numerical resolution.
The shocked AGB wind moves
along the shell toward the equator, as indicated by the $c_a$ contours.
This motion of the shocked AGB wind depends on the relative 
velocity between the shell and undisturbed AGB wind and thus
will still be important even if the 
undisturbed AGB wind has a radial expansion of 20 \vkm{}.
However, this motion is not able to produce mass accumulation 
at the equator due to high density (thus inertia)
of the shell material near the equator.
In the simulation, no Kelvin-Helmholtz instability is seen along 
the contact discontinuity because the velocity changes
smoothly within the shell and the resolution in our simulation is low.
%
%
The velocity decreases from the shocked fast wind 
in the inner shell to the shocked AGB wind in the outer shell,
indicating that the momentum of the shocked fast wind is not transferred
immediately to the shocked AGB wind due to the thermal pressure in the shell.
The shell velocity, $v_s$,
decreases with latitude from the pole to the equator 
but increases with time.
The velocity at the tip where $\theta=0$\degree{} 
is found to be $v_{so}=138(1+0.00165t_{yr})$ \vkm{}, 
where $t_{yr}$ is the outflow age of the fast wind in units of years. 
At the age of 177 years, $v_{so}$ is 178 \vkm{}.

The fast wind itself cools rapidly as it propagates away from the source due to
adiabatic expansion and radiative cooling.
For the temperature of the fast wind, $T_f$, we have
\begin{equation}
\frac{dT_f}{dr} =  - \frac{4}{3}\frac{T_f}{r} - \frac{2 n \Lambda}{3 k v_f}
\end{equation}
Due to radial expansion, the number density
$n = \frac{\rho_f}{1.4 m_H}(\frac{r_f}{r})^2$.
Thus,
\begin{equation}
\frac{dT_f}{dr} = 
- \frac{4}{3}\frac{T_f}{r}
-\frac{\rho_f \Lambda}{2.1 m_H k v_f}(\frac{r_f}{r})^2 
\end{equation}
Integrating this equation numerically, we find that
the fast wind cools down from 10$^4$ to 10$^2$ K after traveling a distance
of 1.2$\times10^{16}$ cm from the source, consistent with the simulation.
The majority of the shell
has a temperature of about 10$^4$ K at which  
the cooling function decreases rapidly (see \S \ref{sec:method}).
The temperature of the shell is highest at the shell boundaries
near the pole where the material is recently and strongly shocked,
about 10$^5$ K for the shocked AGB wind
and a few 10$^4$ K for the shocked fast wind.
The temperature decreases away from the pole 
as the shell velocity decreases.
The temperature also depends on 
the radiative cooling and thus the density.
Near the source where the density is high,
the shocked AGB wind cools to less than 100 K
\cite[due to the cooling of atomic gas at low temperature in our cooling
function, see][]{Dalgarno1972}.
The dense jet-like structure at the tip
also cools to a very low
temperature compared to its surrounding shell material.

In order to further investigate the shell dynamics, we present
the transverse velocity, $v_R$, and longitudinal velocity, 
$v_z$, of the shell material in Figure \ref{fig:model2_vrz}.
The velocity structures of the newly 
shocked fast wind and the newly shocked AGB wind are different. 
The transverse velocity of the newly shocked fast wind 
decreases with increasing $z$. It becomes negative  
near the pole but drops rapidly to zero at the tip,
indicating that the newly shocked fast wind is 
focused toward the pole and shocked strongly at the tip.
At the tip, the sound speed of the newly shocked fast wind, 
$c_s$, is 12 \vkm{}. According to Bernoulli's theorem,
the transverse velocity of the newly shocked fast wind
is limited to $\sqrt{3} c_s$ or 20 \vkm{} for $\gamma=5/3$.
On the other hand, as we go from the tip to the central
source (decreasing $z$),
the transverse velocity of the newly shocked AGB 
wind first shows a rapid increase in the tip 
(indicating that the newly shocked AGB wind at the tip
is squirted sideways along the shell) and then
shows a steady linear decrease.
Since the momentum of the fast wind is mainly along the z-axis,
the longitudinal velocity of the newly shocked AGB wind is very low 
except near the tip. 

In the following, we compare the shell structure and kinematics in the
simulation of model 1 to
a simple analytical momentum-driven shell model given by
\citet{Lee2001}, which
was found to be able to describe the shell dynamics
in similar but isothermal wind simulations.
In this analytical model, 
the thermal pressure in the shell and the motion of the
shocked fast and AGB winds along the shell are not considered.
In this case, the ram pressure of the fast wind is exactly
balanced by that of the AGB wind, so that the shell expands outward
with a constant velocity along any radial direction.
Hence, the shell velocity (see Figure \ref{fig:coor}) is,
\begin{equation}
{\bf v}_s(\theta) = \frac{{\bf v}_f(\theta)}{1+\eta^{-1/2}}
\label{eq:vso}
\end{equation}
and the shell radius is
\begin{equation}
r_s(\theta) = v_s(\theta) t +r_f
\label{eq:rs}
\end{equation}
where $\eta\equiv\frac{r_f^2 \rho_{f}}{r^2 \rho_a} 
= \frac{1}{K}\frac{\dot{M}_{f}}{\dot{M}_{a}}\frac{v_a}{v_{f}}$.
Therefore, the shell velocity at the tip, $v_{so}$, 
is expected to be 138 \vkm{}, 
in agreement with the initial shell velocity as the fast
wind first impacts the AGB wind in the simulation.
However, in the simulation, the shell velocity 
increases with time as the shocked fast and AGB
winds flow along the shell.
As a result, the predicted shell structure
(the dashed line in Figure \ref{fig:model2}a) 
is smaller than that in the simulation. 

The velocity of the newly shocked fast wind 
can be derived from the shell velocity.
However, since the shell velocity increases with time
in the simulation, a mean (averaged over time)
shell velocity, $\bar{v}_s$, is used to derive 
the velocity of the newly shocked fast wind. 
The mean shell velocity is $\bar{v}_s(\theta)=(r_s(\theta)-r_f)/t$.
As expected, its $R$ and $z$ components 
(the solid lines in Figure \ref{fig:model2_vrz})
lie between those of the newly shocked fast
wind and the newly shocked AGB wind.
The velocity of the newly shocked fast wind is
\begin{equation}
{\bf v}_{sf}  = {\bf v}_f-\frac{2(v_f-\bar{v}_s)}{\gamma+1}\cos\alpha 
[\sin(\alpha+\theta) \hat{\bf R} + \cos(\alpha+\theta) \hat{\bf z}]
\label{eq:vsf}
\end{equation}
where $\alpha$ is the angle between the shell normal and the mean shell
velocity (see Figure \ref{fig:coor}) -- 
note that the dependence of various parameters on 
$\theta$ is not explicitly shown in the equation.
Its $R$ and $z$ components (the dashed lines in Figure
\ref{fig:model2_vrz}) are found to match reasonably well 
those of the newly shocked fast wind in the simulation, 
except near the tip where the shocked fast wind shocks with itself,
producing a thermal pressure gradient that
reduces the transverse velocity near the tip.
With this equation, the transverse velocity of the newly shocked fast wind
is found to become negative as $\theta < \theta_c$, where
\begin{equation}
\sin \theta_c = \frac{2(v_f-\bar{v}_s)}{v_f(\gamma+1)}\cos\alpha
\sin(\alpha+\theta_c)
\end{equation}
As a result, the shocked fast wind with  $\theta < \theta_c$
or $3.4$\degree{}
is focused toward the pole, producing a focusing effect that 
produces a shell longer than the momentum-driven shell in the simple
analytical model.

\subsection{Model 2: Effects of changing the opening angle}
The effects of the opening angle 
of the fast wind can be studied by comparing the simulation 
of model 2 to that of model 1.
The simulation of model 2 is presented in Figure \ref{fig:model11}.
The age is 188 years.
In this model, $\dot{M}_f=1\times10^{-5}$ \smyr{},
$v_{fo}=300$ \vkm{}, and $\theta_f=20$\degree{}.
The mass-loss rate is adjusted to give the same $\rho_{fo}$ and thus
the same initial shell velocity at the tip as model 1 
(see Equation \ref{eq:vso}).
The shell dynamics in this model is qualitatively
the same as that of model 1.
However, driven by a fast wind with twice opening angle, the shell has
a length to width ratio of about 3, half of that in model 1.
Hence, under our assumption of the fast wind,
the collimation (length to width ratio)
of the shell increases roughly linearly
with decreasing opening angle of the fast wind.
In addition, since the shell velocity decreases slower away from the pole,
the temperature at the shell boundaries also decreases slower
away from the pole.
Moreover, since the angle between the shell normal 
and the shell velocity is everywhere smaller,
the focusing effect of the shocked fast wind is less.
As a result, the shell velocity increases with time slower
than that of model 1, so that
the shell structure can be described
by the simple momentum-driven shell model
(see the dashed line in Figure \ref{fig:model11}a). 
The shell velocity at the tip is found 
to be $v_{so}=138(1+0.0004t_{yr})$ \vkm{}, 
or 148 \vkm{} at the age of 188 years.

\subsection{Model 3: Effects of changing the velocity}
The effects of the velocity
of the fast wind can be studied by comparing the simulation 
of model 3 to that of model 1.
The simulation of model 3 is
presented in Figure \ref{fig:model15}.
The age is 180 years. 
In this model, $\dot{M}_f=2.5\times10^{-7}$ \smyr{},
$v_{fo}=1000$ \vkm{}, and $\theta_f=10$\degree{}.
The mass-loss rate is chosen to give an initial shell velocity 
of about 150 \vkm{} at the tip, similar to that of model 1 
(see Equation \ref{eq:vso}).
Although the opening angle of the fast wind
is the same, the shell is broader than that of model 1. 
Unlike model 1, the shell is mainly composed of the shocked AGB wind,
as indicated by the $c$ contours.
The shocked fast wind does not cool fast enough and expands
into a low-density but hot, high-pressure and turbulent cocoon
surrounding the unshocked fast wind, as indicated by the
$c_w$ contours.
Being compressed by this high-pressure cocoon,
the fast wind itself becomes a ripple jet
surrounded by a dense sheath of compressed fast wind,
similar to that seen by e.g., \cite{Frank1996a}.
The jet interacts with the AGB wind, 
producing a rippled shell with a convex curvature at the tip.
The shocked AGB wind in the
convex curvature moves toward the symmetry axis.
However, since the shocked AGB wind cannot
flow across the symmetry axis in the simulation, some of it is 
deflected in a backward direction toward the source, 
producing a linear and dense 
structure along the axis inside the shell, as seen in other simulations
\citep{Blondin1993,Dwarkadas1998}. 
As a result, this linear and dense 
structure is likely an artifact
produced by the numerical effect of the simulation.
Since the fast wind itself is compressed by the shocked fast wind and bent 
toward the z-axis, the density of the fast wind 
does not decrease with distance as fast as the density of the AGB wind, 
so that the shell velocity at the tip increases with time.
At the tip, the shell velocity is about 220 \vkm{} and
the temperature of the shell is about 10$^6$ K,
much higher than that in model 1.
On the other hand,
the temperature of the shocked fast wind is a few$\times$10$^7$ K
at the head of the fast wind and 
decreases to a few$\times$10$^6$ K as the shocked fast wind expands 
into the cocoon.
Since the shocked fast wind expands sideways at
the head, there is no focusing effect in this simulation.

The transverse velocity, $v_R$, and
longitudinal velocity, $v_z$, of the shell material are presented in
Figure \ref{fig:model15_vrz}.
Since we are interested in the shell dynamics,
the shocked fast wind in the cocoon 
is excluded in this figure.
The negative $v_R$ is associated with the convex curvature at the tip. 
To avoid confusion, the convex curvature is not included
in the figure of $v_z$.
Unlike model 1, the transverse and longitudinal velocities
both show a convex spur structure with the highest velocity at the
tip, indicating that the shocked AGB wind at the tip,
due to its high thermal pressure and
the thermal expansion of the shocked fast wind at the head,
is squirted sideways along the shell.
This is similar to that seen in the simulations 
of a cylindrical jet \cite[see e.g.,][]{Lee2001}.

In the following, we compare the shell structure and kinematics
in the simulation of model 3
to a simple analytical ballistic bow shock model in \citet{Ostriker2001}, 
which was found to be able to describe the shell dynamics in the simulations
of a cylindrical jet. 
In the ballistic bow shock model,
the shocked AGB wind in the shell is assumed to be accelerated by 
an initial impulse at the head --
the thermal pressure forces of 
the shocked fast wind and shocked AGB wind
are ignored subsequent to the initial impulse.
At the tip, the shell velocity, $v_{so}$, is about 220 \vkm{} and
the sound speed is about 65 \vkm{} in the simulation.
Therefore, $v_R$ is limited to the smaller value of 
$\sqrt{3}c_s$ and $v_{so}/2$, or
110 \vkm{}, consistent with the ballistic bow shock model.
Using  Equations 18-22 in \citet{Ostriker2001} and
assuming a mean jet radius of $R_j=4\times10^{15}$ cm 
for the ripple jet (see Figure \ref{fig:model15}) and $\beta=4$ 
\cite[a typical value used for the initial impulse at the head, 
see][]{Lee2001,Ostriker2001},
we calculated the shape of
the shell (the dashed line in Figure \ref{fig:model15}), 
the mean shell velocity averaged over the shell thickness
(the solid lines in Figure \ref{fig:model15_vrz}), and the immediate
postshock velocity of the AGB wind
(the dashed lines in Figure \ref{fig:model15_vrz}).
As can be seen, the shape and the transverse 
velocity of the shell in the simulation
can be reasonably reproduced with the ballistic bow shock model. 
However, the simple ballistic bow shock model
cannot reproduce the longitudinal velocity that well. 
In the simulation,
the material in the shell consists of two components, 
one with a velocity close to the immediate postshock velocity
and the other with a velocity close to the mean shell velocity.


\subsection{Model 4: Effects of the time variations}
The effects of the time variation
of the fast wind can be studied by comparing the simulation 
of model 4 to that of model 1.
The simulation of model 4
is presented in Figure \ref{fig:model2P}. The age is 166 years.
In this model, the fast wind has the same parameters as that in model 1 
but has a time variation in the density and velocity
with $A=0.5$ and $P=22$ years 
(see Equation \ref{eq:ft}).
The structure and velocity of the shell in this model 
are similar to those of model 1. 
There is also a jet-like structure at the tip.
Inside the shell, however, there are periodic
internal shock pairs (i.e. forward and backward shocks), 
which are clearly seen
in the temperature distribution, 
propagating away from the source at the velocity $v_f$. 
The wings of the internal shocks grow with time and eventually 
impact the inner surface of the shell. 
The curvature of the internal shocks are
determined by the angular distribution of the fast wind's velocity and thus
well described by (dashed lines in Fig. \ref{fig:model2P}a)
\begin{equation}
r (\theta) = v_f (\theta) t_i + r_f
\label{eq:internal}
\end{equation}
where $t_i$ is the age of each internal shock.
The thickness of the internal shock
pairs (i.e the separation between the forward and backward shocks)
increases with the distance due to thermal expansion.
The thickness along the $z$-axis ($\theta=0$\degree) is found to be
\begin{equation}
\tau=190+\frac{c_s}{v_f}(z-r_f)\; \textrm{AU}
\end{equation}
where $c_s = 11$ \vkm{}, the sound speed 
at the shock boundaries in the simulation.
The density of the internal shocks decreases with the distance 
due to radial expansion. 
The shocked fast wind also moves along the internal shock surfaces,
as indicated by the $c_w$ contours.
The leading and the internal shocks
all show highest temperature at the boundaries 
where the material is newly shocked. 

\subsection{Shell dynamics}

It is clear that the velocity of the fast wind is
a crucial parameter in
determining the shell dynamics, e.g., whether the shell is momentum-driven
or driven by thermal pressure.
This was recognized by \citet{Dwarkadas1998} who carried out simulations of
an ISW model in which the fast wind velocity ramped up from 25 to 1000
\vkm{} -- their simulated nebulae evolve through an initial momentum-conserving phase before
entering the energy-conserving stage.
In our simulations, the parameters 
are chosen to give a shell velocity
of about 150 \vkm{} at the tip,
comparable to that seen in CRL 618.
For a given shell velocity at the tip, $v_{so}$,
the velocities of the shocked AGB wind and shocked fast 
wind at the tip are $v_{so}$ and $v_{fo}-v_{so}\equiv v_{sfo}$, respectively.
In order to balance the ram pressures of the shocked fast
wind and shocked AGB wind at the tip, the density of the fast wind 
along the z-axis ($\theta=0$\degree) is required to be 
\begin{equation}
\rho_{fo}=\frac{\dot{M}_{a}}{4 \pi r_{f}^2 v_a}
(\frac{v_{so}}{v_{sfo}})^2
\label{eq:rf}
\end{equation}
giving a kinetic energy injection rate along the z-axis of
\begin{equation}
L_{ko}=
\frac{K}{2}\frac{\dot{M}_{a}v_{fo}^3}{v_a}(\frac{v_{so}}{v_{sfo}})^2
\label{eq:Lk}
\end{equation}
For a 300 \vkm{} fast wind, the shocked AGB wind and shocked
fast wind both have a velocity of about 150 \vkm{} at the tip.
The cooling is fast enough to remove the heat obtained 
from the fast wind,
so that the shell is thin and momentum-driven. 
For a 1000 \vkm{} fast wind, however,
the shocked fast wind has a velocity of 850 \vkm{} at the tip.
Thus, the fast wind is required to be more tenuous 
than that of 300 \vkm{} fast wind
(see Eq. \ref{eq:rf}). However, the energy injection rate is about the same
(see Eq. \ref{eq:Lk}), so that the shocked fast wind 
does not cool fast enough and 
expands into a hot cocoon.
The fast wind itself becomes a ripple jet surrounded
by a dense sheath of compressed fast wind.
As a result, the shell is essentially driven by a jet.  
At a velocity of about 150 \vkm{}, 
the shell is radiative and its kinematics can be described 
by a ballistic bow shock model.

\subsection{Comparison with previous analytical models}
Recently, 
\citet{Soker2002} has also developed analytical models
of highly CFWs in order to reproduce highly 
collimated shells in PPNs and PNs.
He defines two cases: 
(i) the ``slow-propagating" jet, in which the shell
at the tip proceeds at a speed $v_{so} \ll v_{fo}$, which can be compared
to our simulations of model 3;
and (ii) the ``fast-propagating" jet, 
in which $v_{so} \sim v_{fo}$. In case (ii), 
the density of the gas in the jet is much larger than that of the AGB wind;
this condition is not met in any of our simulations.
In Soker's case (i) above, 
the shocked fast wind is assumed to be adiabatic and expands sideways 
behind the shell of the shocked AGB wind, similar to that seen in our
simulation of model 3.
He also assumes that the shell is continuously accelerated 
by the thermal pressure force of the shocked fast wind and
expands laterally perpendicular to the outflow axis.
In our simulation, however, the ram pressure in the shell is much greater
than the thermal pressure of the shocked fast wind except at the head.
As a result, the force due to the thermal pressure of the shocked fast wind
can be ignored subsequent to an initial impulse at the head, so that
the shell dynamics is in the ballistic limit.
In addition, unlike his model, 
the shell in our simulation also expands significantly in the $z$ direction 
(see Figure \ref{fig:model15_vrz})
as the momentum of the fast wind is 
transferred to the shocked AGB wind.
As a result, his model predicts a lobe much wider than that in our
simulation (model 3).

\section{Comparison with observations}
In the following, we compare our simulations to 
the observations of CRL 618.
CRL 618 is a young PPN located at a distance of 0.9 kpc
\citep{Goodrich1991}. It
shows several narrow
lobes at different orientations in HST images 
\citep{Trammell2000}, perhaps resulting from multiple
ejections at different orientations. 
The general structures of the different lobes 
are similar. We compare our simulations 
to the northwestern (W1) lobe, which is separated from other 
lobes and shows clear position-velocity diagrams 
in the groundbased observations.
This lobe has an inclination of about 30\degree{} 
into the plane of the sky (SCSG02) and a length of about 10$^{17}$ cm.
Since molecular cooling is not included in our simulations,
we only compare the optical emission.
In the simulations, the emission is assumed to be
optically thin and arising from gas in local
thermal equilibrium (LTE) (see Appendix).
We assume a distance of 1 kpc and 
an inclination of 30\degree{}, the values
similar to those of CRL 618.

\subsection{Morphology}\label{subsec:morphology}
The W1 lobe of CRL 618 is 
highly collimated with a length to width ratio of about 7. 
In optical emission,
several bow-like (or ripple-like) structures are seen
within the body of the lobe (see Figure \ref{fig:crl618}).
The \SIIratio{} ratio 
is about 0.5 and roughly constant in the lobe (SCSG02).
We compare our simulations to these features 
in order to determine which model can better produce the
morphology of the lobe. 

With a highly collimated fast wind, model 1 is able to
produce a narrow lobe similar to the W1 lobe.
In this model, the velocity of the shocked AGB wind is higher than 
that of the shocked fast wind, so that
the optical emission arises mostly 
from the newly and strongly shocked AGB wind,
showing a single bow-like structure at the tip 
(see Fig. \ref{fig:model2_emis}).
From [OIII] to [NII] to [SII] to [OI],
the emission traces the material at lower and lower temperature,
showing more and more extended bow-like emission structure.
At the tip, the electron density is about 10$^4$ \cm3{} and
the \SIIratio{} ratio is about 0.5, similar to that seen in the W1 lobe.
The emission from the newly shocked fast wind
could be as strong as that from the newly shocked AGB wind, 
showing two bow-like structures at the tip, as that seen in model 2 
(see Fig. \ref{fig:model11_emis}), in which 
the velocities of the shocked AGB wind and shocked fast wind happen to be 
comparable, about 150 \vkm{}.
Due to its high density, the material between the two bow-like structures 
cools down quickly and radiates very weakly in optical emission.
As a result, a 300 \vkm{} fast wind with a small opening angle
can produce one or two close bow-like 
structures at the tip similar to that seen in the W1 lobe,
however, it has difficulty producing other bow-like
structures in the W1 lobe.

On the other hand, 
a model with a 1000 \vkm{} fast wind
has difficulty producing a narrow lobe similar to that of the W1 lobe.
Although it has the same opening angle of the fast wind 
as that in model 1,
model 3 produces a lobe much broader than the W1 lobe.
We have performed a simulation using a fast wind with
an opening angle half of that in model 3 but with the same velocity and
density along the axis, 
however, the lobe in that simulation is only 15 percent smaller than that of
model 3, and thus still much broader than the W1 lobe.
As a result, due to thermal expansion of the shocked fast wind,
the collimation of the lobe produced by a high speed fast wind
increases slowly with decreasing opening angle of the fast wind.
In model 3, the optical emission shows a bright
bow-like structure at the tip (see Fig. \ref{fig:model15T2_emis}).
The emission becomes weak away from the tip. 
Except near the tip,
the [OIII] emission is from the inner shell while
the [OI] emission is from the outer shell.
The \SIIratio{} ratio
is everywhere about 0.5, similar to that seen in the W1 lobe. 
There are ring-like emission structures arising 
from the ripple in the shell in the simulation
(see Figure \ref{fig:model15}).
However, these ring-like structures are much more rounded than
the observed bow-like structures in the W1 lobe, suggesting that
the bow-like structures
are not associated with the ripple in the shell.
There is also a cone-like structure along the axis associated 
with the dense sheath of compressed fast wind around the jet (see Fig.
\ref{fig:model15}).
Since this structure is not seen in the W1 lobe,
the hot and high-pressure cocoon that
produces the dense sheath of compressed fast wind is probably not 
present in the W1 lobe.
As a result, we conclude that the velocity of the shocked fast wind in the W1 lobe is
probably smaller than that in model 3.
Since the hot and high-pressure cocoon radiates in X-ray emission,
future X-ray observations will be able to
check the existence of the hot cocoon in the W1 lobe.
For future comparison,
we present the
X-ray emission and spectrum in the 0.2 -1.5 keV energy range derived from
model 3 in Figure \ref{fig:model15T2_xray}.
The total X-ray luminosity is $L_x=1.5\times10^{29}$ ergs s$^{-1}$.
The X-ray is strongest at the head where the shock is the strongest.
There are strong emission lines from oxygen and iron in the spectrum.
Current upper limits on the X-ray flux from CRL618 are roughly
a factor 10 larger
than our model flux. CRL618 was not detected in a 18697 sec observation
with the HRI instrument onboard ROSAT \citep{Guerrero2000}. If we assume
that a minimum of 9 photons are required for a 3$\sigma$ detection, we can
use the detection of BD+30\degree{}3639 with HRI (172+/-13 counts in 8740
sec, Leahy et al 1998) and its measured X-ray flux 5.7$\times 10^{-13}$ 
erg cm$^{-2}$ s$^{-1}$ to
put a roughly upper limit on the X-ray flux of CRL618 of $<$1.4$\times 10^{-14}$
erg cm$^{-2}$ s$^{-1}$ (1.7$\times10^{30}$ erg s$^{-1}$).


The series of optical bow-like structures observed 
in the body of the W1 lobe (as well as other lobes) but not reproduced
in the previous models, suggest that 
the fast wind may have time variations in density and velocity.
With an episodic fast wind, model 4 produces a series of optical 
bow-like structures in the body of the lobe (see Fig. \ref{fig:model2P_emis}).
The two bright bow-like structures (B$_1$ and B$_2$) at the tip trace 
the newly shocked AGB wind in the outer shell and the newly
shocked fast wind in the inner shell impacted by the internal shocks.
However, in contrast to the observations that show  
a series of bow-like structures of similar intensity within the lobe, 
the optical emission of our model internal
bow-like structures becomes progressively weaker
away from the source (due to radial expansion and radiative cooling).
The results for the \SIIratio{} ratio
are also inconsistent with the observations of
the W1 lobe, which show a roughly constant value of 
0.5 with fluctuations from 0.4 to 0.6.
In our model, this ratio is (a) 
about 1.5 for the shell along the body of the lobe,
(b) 0.6 in the bow-like structure near the source 
(and increases away from the source).
Although there is also emission 
from the inner shell (that is impacted by the wings of 
the internal shocks), this emission does not have
the observed bow-shaped geometry.

One way to resolve the above discrepancies between our model 4 
and the observations
is to have a mass-loss rate of the fast wind ($\dot{M}_f$)
that decreases with time.
However, in order 
to produce adequate optical emission along the body of the lobe,
$\dot{M}_f$ at time $t=0$ would have to be a factor of $\sim$
10 larger.
Another possibility is for the underlying driving agent to be
an episodic {\it cylindrical} jet, which has a density constant 
with the distance. 
Simulations with episodic
cylindrical jets have been performed by a number of authors 
in order to reproduce the morphology and kinematics of protostellar outflows 
\citep{Stone1993b,Raga1993,Biro1994,Suttner1997,Lee2001}. 
Such jets can also produce a series of
internal bow-like structures along the body of the lobe.
Without radial expansion, the emission of
these structures does not decrease as fast as that in model 4.
In addition, each of these structures
has overall stronger emission than its counterpart in our model 4
because it closes back to, and is shocked by, the one behind it
\cite[see Fig. 6 and 7 in][]{Biro1994}.

\subsection{Kinematics}

The position-velocity (PV) diagrams of optical emission for
cuts along the axis of the W1 lobe 
show a nonlinear increase in velocity with 
increasing distance from the source and 
a broad range of velocities at the tip with a FWZI up to 200 \vkm{} 
(SCSG02).
In this section, we present the PV
diagrams of the [OI]$\lambda6300$ emission,
the most extended emission, 
derived from our simulations 
and compare them to that of the W1 lobe.
The diagrams are also degraded to low resolutions for comparing with the
current observations.

The PV diagrams derived from the simulation of model 1 are
presented in Figure \ref{fig:model2_OI}.
At an inclination $i=30$\degree{},
the newly shocked fast wind is 
projected at a velocity higher than that of the 
newly shocked AGB wind.
At low resolution, the diagram shows 
two elliptical structures attached together.
These diagrams show a broad range of velocities at the tip, 
qualitatively consistent with that seen in the W1 lobe.
However,
since the emission is mostly from the newly shocked AGB wind, 
the velocity range is about $v_{so} \sin i$ = 89 \vkm{},
much smaller than that seen in the W1 lobe.
In order to produce the observed velocity range,
the shock velocity at the tip would have to be 400 \vkm{},
resulting in a very high temperature for the shell ($> 10^6$ K)
As mentioned, 
the emission from the newly shocked fast wind could
be as strong as that from the newly shocked AGB wind.
In that case, the detailed PV structure would be similar to 
that of model 2 (see Fig. \ref{fig:model11_OI}) and
the velocity range at the tip would be 
about $v_{f} \sin i$ = 150 \vkm{}, similar to that seen in the W1 lobe. 

The PV diagrams for model 3 are presented in Figure \ref{fig:model15T2_OI}.
They show a convex spur structure on the redshifted side, 
i.e., a nonlinear increase in velocity toward the tip 
with increasing distance from the source.
There is also emission on the blueshifted side.
A broad range of velocities is also seen at the tip.
The maximum redshifted velocity is $v_{so}-\frac{v_{so}}{2}\sin
i = 165$ \vkm{} while the maximum blueshifted velocity is 
$-\frac{v_{so}}{2}\sin i = -55$ \vkm{}, so that the velocity range at the tip is
equal to the shock velocity at the tip independent of the inclination, 
as found in a jet-driven bow shock model by \citet{Hartigan1987}.
As a result, this model produces a similar velocity range
as that seen in the W1 lobe.

The PV diagrams for model 4 are presented in Figure \ref{fig:model2P_OI}.
The PV structures associated with the leading shock 
are similar to that seen in model 1.
There are also structures associated with the internal shocks
projected at a velocity higher than that of the leading shock,
showing a series of increases in velocity.
These PV diagrams are different from that of the W1 lobe,
which shows a single nonlinear increase in velocity 
with increasing distance.
Hence, the bow-like structures seen in the W1 lobe
may not be associated with internal shocks.

\subsection{Line Ratios and Intensities}
Here we compare the optical emission line fluxes
derived from our simulations to the observations (Table \ref{tab:emis}).
Since models 1 and 4 can produce a lobe with a collimation similar to that
of the W1 lobe, we compare these in detail to the observations.
The fluxes always peak at the tip of the lobe in 
both the simulations and observations. 
However, the ratios of the [OI] to [SII] peak fluxes derived from our
simulations are too low compared to those derived from high resolution HST
observations, indicating that 
the temperatures of the shocked AGB wind at the tip in 
our simulations ($\sim 2\times 10^4$ K) 
are always much higher than observed ($\sim 8\times 10^3$ K) (see Fig.
\ref{fig:ratio}). 
This conclusion is further supported by the groundbased observations.
For example, the ratios of the [OIII] and [NII] total fluxes
to the [SII] total fluxes in the tip of the lobe
in the simulations are significantly
higher than observed, which also indicates that the temperatures 
at the tip in the simulations are much higher than observed.
Notice that in the observations,
the  [OIII] to [SII] ratio 
indicates a significantly higher temperature
($\sim$ 30,000 K) than that indicated by 
the ratios of [NII] and [OI] to [SII] ($\sim$ 15,000 K).
Thus, the [OIII] emission comes from a region hotter than that of the
[NII] and [OI] emission, consistent with an existence of 
a temperature stratification in the tip. Our models also show
similar temperature stratification, although as mentioned above,
the model temperatures are higher than observed.

The high temperatures at the tip of the shell in our simulations
are produced by
the dense and cold jet-like structure at the tip
as it outruns the shell and shocks the AGB wind at the head 
before the shell does. 
The shell at the tip then interacts with the shocked AGB wind, resulting in
a temperature higher than observed.
The jet-like structure is formed as the shocked fast wind material flows
toward the tip \citep{Canto1988,Frank1996b}.
In reality, the cold dense jet-like material may mix with the hot
shell material at the tip of the lobe (e.g., due to instabilities), 
significantly lowering the temperature there.
Further simulations with high resolution are needed to test this idea.
It is interesting to note that 
the observed line ratios at the tip of the W1 lobe 
(and thus the inferred temperature) are similar to those
in the internal shocks in model 4. These internal shocks lack 
the jet-like structure at the tip in our simulations, thus supporting the idea
that the jet-like structure is responsible for the high temperatures at the
tip.

Although these models (1 and 4)
produce a peak [SII] flux
comparable to that of the observations,
the total [SII] fluxes in our simulations are much less than observed
because of the lack of significant emission within the body of the lobe.
New simulations will need to address this discrepancy.
Note that such simulations will also have to explain that
the observed ratio of the [OI] to [SII] fluxes in the body of the lobe
is smaller than that at the tip (which 
implies that the body of the lobe is hotter than the tip).


\subsection{The nature and origin of the collimated fast wind in CRL 618}

The fast wind in CRL 618 is very unlikely to be a steady wind.
This conclusion is most directly supported by the presence 
of the optical bow-like structures within the body of the lobe 
(see discussion in \S \ref{subsec:morphology}).
Furthermore, CRL 618 is known to have a fast ($\sim$ 200 \vkm{}), 
compact (size $\lesssim$ 5") bipolar molecular outflow \citep{Neri1992}.
Only our pulsed wind model produces material moving at such high speeds
close to the central source.

Since AGB winds are thought to be driven by radiation pressure on dust
grains \cite[see review by][]{Habing1996}, 
we need to consider the possibility that the fast winds in PPNs 
are also powered by radiation pressure.
In our models, 
the momentum flux per steradian of the fast wind is
\begin{equation}
\dot{P}_f = \frac{\dot{M}_f v_{fo}}{4 \pi K}
\exp[-3(\frac{\theta}{\theta_f})^2] \equiv \dot{P}_{fo}
\exp[-3(\frac{\theta}{\theta_f})^2] 
\end{equation}
The luminosity of CRL 618 is $L\sim 10^4$\sl{} \citep{Goodrich1991},
giving a radiation momentum flux per steradian of 
$\dot{P}_L =L/4\pi c \sim 10^{34}$ erg cm s$^{-1}$ yr$^{-1}$ sr$^{-1}$.
Thus, $\dot{P}_f \gg \dot{P}_L$ within the opening angle of the CFW in our
models (see Table \ref{tab:parameters}), implying that
the CFW can not be driven by radiation pressure.
It is likely that the fast wind is launched by magneto-centrifugal forces
from a magnetized accretion disk and star system \citep{Blackman2001}.
The accretion disk may be formed in a binary system as the material flows
from one star to another \citep{Soker2000}.
In this case, the gravitational binding energy in the accreting material is
converted into the kinetic energy of the fast wind.

\subsection{Application to other PPN: OH 231.8+4.2}
The simulation of model 2 may provide
a clue to the formation of another PPN, OH 231.8+4.2.
This PPN is bipolar, showing two strong and 
wide bow-like structures in H$\alpha$ at the tip 
of both the northern and southern lobes \citep{Bujarrabal2002}.
A fast, collimated, dense CO structure is seen inside each lobe 
along the axis \citep{Sanchez2000}.
Recently, it has been proposed that
the two bow-like structures in both northern and southern
lobes trace the forward and backward shocks 
driven by a dense jet \citep{Bujarrabal2002}. 
Models with an overdense cylindrical 
jet have difficulty producing as wide and strong bow-like emission
for the backward shock as that for the forward shock due to the fast cooling
of the backward shock \citep{Raga1995}.
In contrast, a fast collimated wind can readily produce two
strong and wide bow-like structures at the tip 
as those seen in model 2 (see Figure \ref{fig:model11_OI}).
Coincidentally, the collimation of the lobe and
the velocities of both the forward and backward shocks in this model are
similar to those derived by \citet{Bujarrabal2002} for
the northern lobe of OH 231.8+4.2.
As a result, if the two bow-like structures are to be associated with 
the forward and backward shocks, 
a tenuous collimated fast wind may be needed.
If this is the case, the observed dense CO collimated structure 
would represent the dense core of a cool, latitudinally stratified fast wind.
The lengths of the two lobes are different and may 
be due to an intrinsic difference in the velocity on each side of the
bipolar fast wind \citep{Soker2002}. 



\section{Summary and Conclusions}

We have presented a number of simulations
of a collimated fast wind interacting with
a spherical AGB wind. 
In our simulations, we have used a dense AGB wind with a high mass-loss rate
typical of PPNs like CRL 618. The parameters of the fast wind
are chosen to give a shell velocity at the tip
of about 150 \vkm{},
comparable to that seen in CRL 618. Under these assumptions,
the main results from the simulations are the following:
\begin{enumerate}
\item 
The shell dynamics is determined by
the velocity of the fast wind.
For a 300 \vkm{} fast wind, 
the cooling of the shocked fast wind is fast enough to 
remove the heat obtained from the fast wind,
so that the shell is thin and momentum-driven. 
For a 1000 \vkm{} fast wind, however,
the shocked fast wind does not cool fast enough and 
expands sideways into a hot cocoon surrounding the fast wind.
The fast wind itself becomes a ripple jet,
so that the shell is essentially driven by a jet  
and its kinematics can be described 
by a ballistic bow shock model.
\item 
Although the AGB wind is spherical, 
the shell driven by a collimated 
fast wind has the shape of a highly collimated lobe.
For a shell driven by a 300 \vkm{} fast wind,
the collimation of the shell increases roughly linearly with the decreasing
opening angle of the fast wind. 
However, for a shell driven by a 1000 \vkm{} fast wind, 
the collimation increases slowly with the decreasing opening angle.
\item
A time-varying velocity of the fast wind
produces a series of internal shock pairs
interacting with the inner surface of the shell. 
Due to radial expansion of the fast wind,
the density of the internal shocks decreases rapidly with distance.
\end{enumerate}

We have also derived various emission diagnostics from our simulations.
The main results are the following:
\begin{enumerate}
\item 
For a 300 \vkm{} fast wind, 
the shell is composed of the shocked AGB wind (outer shell)
and the shocked fast wind (inner shell). 
The optical emission arises mostly from the newly shocked AGB wind at the
tip of the lobe, forming a bow-like structure. 
If the velocities of the 
shocked fast wind and the shocked AGB wind are similar, 
two bow-like structures are seen at the tip because
the newly shocked fast wind also contributes significantly to
the emission.
\item
For a 1000 \vkm{} fast wind, the shell is dominantly
composed of the shocked AGB wind, and
the optical emission again forms a
bow-like structure at the tip. The shocked fast wind, however,
forms a tenuous and hot, X-ray emitting structure interior to the shell.
\item
When the velocity of the 
fast wind is time-varying,
bow-shaped optical emission structures are seen associated with the
internal shocks along the body of the lobe. 
This emission decreases rapidly away from the source.
\item
The position-velocity (PV) diagrams from all our models
show a broad range of velocities at the tip.
However, the velocity range
at the tip depends on the shell dynamics and the relative contributions of
the shocked fast wind and the shocked AGB wind.
For a momentum-driven shell, the velocity range depends on the inclination.
For a shell driven by a ballistic bow shock, however,
the velocity range is independent of the
inclination and is equal to the shock velocity at the tip.
When the velocity of the fast wind is time-variable,
a series of increases in the velocity are seen in the PV diagrams.
\end{enumerate}

Comparing our simulations to the observations of the PPN CRL 618,
we find that 
\begin{enumerate}
\item
A 300 \vkm{} collimated fast wind 
with an opening angle of 10\degree{}
can readily produce a highly collimated lobe similar 
to the W1 lobe in this object.
However, a 1000 \vkm{} fast wind, 
even with an opening of 5\degree{},
has difficulty producing such a highly collimated lobe.
\item
Our models can produce an optical bow-like
structure at the tip similar to that seen in the W1 lobe.
However, they have difficulty producing several additional 
bright optical bow-like structures seen in the body of the lobe.
A fast wind with a less steep (than $r^{-2}$)
radial density gradient may help to remove this difficulty.
\item
The PV diagrams derived from our simulations are
qualitatively consistent with the observations.
However, the observed velocity range at the tip
of the W1 lobe is larger than the momentum-driven shell model and
may favor a shell driven by a ballistic bow shock.
\item
The \SIIratio{} ratios at the tip of the lobe
in our simulations are all about 0.5 
(implying an electron density of $\sim 10^4$ \cm3{}), as observed.
\item
Specific line ratios indicate that the temperatures at the tip of the lobe
in our simulations are higher than observed. 
This may result from insufficient mixing of hot and cold material 
in the tip in our simulations. A comparison of the inferred temperatures
from the [OIII]/[SII] ratio to that from the [NII]/[SII] and [OI]/[SII]
ratios indicates a temperature stratification in the tip, both for the
simulations and observations.
\item
The collimated fast wind is unlikely to be steady and is not radiatively
driven.
\end{enumerate}



\acknowledgements
We thank Noam Soker, Vikram Dwarkadas and Garrelt Mellema for their comments 
about this paper.
We thank C. S{\' a}nchez Contreras for fruitful conversations about the
observations of CRL 618 as well as providing unpublished data on line ratios.
This work was performed while C.-F. Lee held a National Research Council
Research Associateship Award at the Jet Propulsion Laboratory, Caltech.
R.S. acknowledges support by NASA through a Long Term Space Astrophysics
grant (no. 399-20-61-00-00).

\begin{appendix}
\section{Emission calculation}
The optical emission is assumed to be a thermal equilibrium (LTE)
optically thin emission, so that the emissivity is given by
\begin{equation}
\epsilon=\frac{x f_i f_u n A_{ul} E_{ul}}{4 \pi}
\end{equation}
where $x$, $f_i$, $f_u$, $A_{ul}$, $E_{ul}$ are
the abundance of the element relative to hydrogen nuclei,
ionization fraction of the ions,
fraction of the ions in the upper excited state,
Einstein A coefficient, and energy separation
between the upper and lower excited states, respectively.
The abundances relative to hydrogen nuclei are assumed to be
4.17$\times10^{-4}$,
9.77$\times10^{-6}$ and
1.28$\times10^{-4}$ for O, S and N, respectively.
The ionization fractions of [OI] and [OIII] as a function of temperature
are taken from
\citet{Nahar1999}, [NII] from \citet{Nahar1997} and
[SII] from \citet{Shull1982}. 
$f_u$ is obtained by solving the level populations
with the relevant atomic quantities taken from \citet{Mendoza1983}.
The X-ray emission is also
assumed to be an equilibrium optically thin emission. It is
calculated with the \citet{Raymond1977} model and a 
solar abundance from \citet{Anders1989}.
\end{appendix}

\newpage

\newpage

%
%
\begin{deluxetable}{lcccccc}
\tabletypesize{\normalsize}
\tablecaption{Parameters For Fast Winds\label{tab:parameters}}
\tablewidth{0pt}   
\tablehead{
\colhead{Model} & \colhead{$\dot{M}_{f}$} & \colhead{$v_{fo}$}
& $\theta_{f}^{a}$ & $R^{b}$ & \colhead{$K$} & 
\colhead{${\dot{P}_{fo}/\dot{P}_L}^{c}$} \\
\colhead{}     & \colhead{(10$^{-7}$ \smyr)}
& \colhead{(100 \vkm)} & \colhead{(degree)} & \colhead{(10$^{-3}$)} 
& (\%) &
}
\startdata
1       & 25    &  3  & 10 & 7.6 & 86.5 &158 \\
2       & 100   &  3  & 20 &30.2 & 39.6 &158 \\
3       & 2.5   & 10  & 10 & 7.6 & 86.5 & 53 \\
4$^{*}$ & 25    &  3  & 10 & 7.6 & 86.5 &158 \\
\enddata
\tablenotetext{a}{Opening angle}
\tablenotetext{b}{Total mass-loss rate within 10\degree{} from the pole/Total mass-loss rate}
\tablenotetext{c}{Momentum flux per steradian of the fast wind at the
pole/Radiation momentum flux per steradian}
\tablenotetext{*}{Model 4 is Model 1 with time variation in 
the density and velocity.}
\end{deluxetable}

\begin{deluxetable}{lccccc}
\tabletypesize{\normalsize}
\tablecaption{Emission line diagnostics\label{tab:emis}}
\tablewidth{0pt}
\tablehead{\multicolumn{6}{c}{Peak Intensity}}
\startdata
&\multicolumn{4}{c}{Model} & Observations$^a$\\
	& 1      &  2    	& 3      & 4  &  \\
\hline
$[$OI$]\lambda$6300\AA$^c$    & 0.5      &  0.9       & 1.3      & 0.53 & 5.4\\
$[$SII$]\lambda\lambda$6730,6716\AA$^b$ & 1.1e-13  &  6.1e-13  & 1.1e-12 & 8.1e-14 & 1.4e-13\\
$[$NII$]\lambda$6583\AA$^c$            & 5.0      &  6.8      & 9.3    & 3.6 & --\\
$[$OIII$]\lambda$5007\AA$^c$            & 5.6      &  7.5      & 19.7    & 90 & -- \\
\\ \\
\multicolumn{6}{c}{Total Flux in the Peak$^d$}\\
\hline
&\multicolumn{4}{c}{Model} & Observations$^e$\\
	& 1      &  2    	& 3      & 4  &  \\
\hline
$[$OI$]\lambda$6300\AA$^g$	& 0.6      &  0.8      & 0.8      & 0.7  & 1.5\\
$[$SII$]\lambda\lambda$6730,6716\AA$^f$  & 6.6e-15  &  3.2e-13  & 3.7e-13 & 5.1e-15 & 1.1e-14\\
$[$NII$]\lambda$6583\AA$^g$             & 3.9      &  5.0      & 6.5    & 2.0 & 1.3\\
$[$OIII$]\lambda$5007\AA$^g$             & 2.1      &  3.2      & 9.8      & 0.7 & 0.07\\
\enddata
\tablenotetext{a}{Adopted from the HST images in Sahai et al. (2003)}
\tablenotetext{b}{Peak Intensity in unit of erg s$^{-1}$ cm$^{-2}$
arcsec$^{-2}$.}
\tablenotetext{c}{Peak Intensity normalized to that of
[SII]$\lambda\lambda$6730,6716\AA}
\tablenotetext{d}{Measured in a 2"$\times2"$ box covering the tip of the W1 lobe.}
\tablenotetext{e}{From groundbased observations (SCSG02 and
 priv. comm.)}
\tablenotetext{f}{Total flux in the peak in unit of erg s$^{-1}$ cm$^{-2}$.}
\tablenotetext{g}{Total flux in the peak normalized to that of
[SII]$\lambda\lambda$6730,6716\AA}

\end{deluxetable}

\clearpage

%
%
\begin{figure} [!hbp]
\centering
\epsscale{0.7}
\epsfbox{\cdir/model2.eps}
\figcaption[]
{Simulation of model 1 
as the fast wind propagates to a distance of
$\sim$ 10$^{17}$ cm, a length similar to that of the lobes in CRL 618.
The age of the fast wind is 177 years. In this model,
$\dot{M}_f=2.5\times10^{-6}$ \smyr{},
$v_{fo}=300$ \vkm{}, and $\theta_f=10$\degree{}.
(\tlabel{a}) shows the contours of the fast wind tracer, $c$,
superposed on the density distribution. Contours are 0.1, 0.5 and 0.9.
The dashed line is the shell structure derived from the momentum-driven shell
model with Equation \ref{eq:rs}.
(\tlabel{b}) and (\tlabel{c}) show the contours of $c_w$ and $c_a$, 
respectively, superposed on the density distribution. 
$c_a$ and $c_w$ trace respectively 
the flows of the AGB wind and the fast wind 
emanating within the angular segment  5\degree{}$<\theta<$7.5\degree{}.
Contours are 0.1, 0.5 and 0.9.
(\tlabel{d}) shows the velocity distribution
superposed on the pressure distribution.
(\tlabel{e}) shows the temperature distribution.
The gray-scale wedges along the right side indicate
the values for the gray-scale images.
(\tlabel{A}), (\tlabel{B}) and (\tlabel{C})
show respectively
the blow-ups for the regions in the boxes \tlabel{A}, \tlabel{B} and
\tlabel{C}. Regions i, ii, iii, iv and v indicate unshocked fast wind,
shocked fast wind, mixture of shocked fast and AGB winds, shocked AGB wind
and unshocked AGB wind, respectively. 
Contact discontinuity is located in region iii.
\label{fig:model2}
}
\end{figure}
\clearp

\begin{figure} [!hbp]
\centering
\epsscale{0.9}   
\epsfbox{\cdir/model2_vrz.eps}
\figcaption[]
{Transverse velocity ($v_R$) and longitudinal velocity ($v_z$) of the shell
material for model 1 at 177 years.
The gray-scale indicates the mass column density on a linear stretch.
The solid lines indicate the $R$ and $z$ components of the mean shell 
velocity obtained from the simulation.
The dashed lines indicate the $R$ and $z$ components of the velocity of the
newly shocked fast wind derived from Equation \ref{eq:vsf}.
\label{fig:model2_vrz}
}
\end{figure}
\clearp

\begin{figure} [!hbp]
\centering
\epsscale{0.6}   
\epsfbox{\cdir/coor.eps}
\figcaption[]
{Coordinate system used for the shell dynamics.
The thick line indicates the shell. 
$v_f$, $v_s$ and $r_s$ are the velocity of the fast wind,
the shell velocity and the shell radius, respectively.
$\alpha$ is the angle between the shell normal
and the shell velocity. 
\label{fig:coor}
}
\end{figure}
\clearp

\begin{figure} [!hbp]
\centering
\epsscale{0.8}   
\epsfbox{\cdir/model11.eps}
\figcaption[]
{Simulation of model 2.
The age is 188 years.
In this model, $\dot{M}_f=1\times10^{-5}$ \smyr{},
$v_{fo}=300$ \vkm{}, and $\theta_f=20$\degree{}.
The captions are the same as Figure
\ref{fig:model2}.
\label{fig:model11}
}
\end{figure}
\clearp

\begin{figure} [!hbp]
\centering
\epsscale{0.8}   
\epsfbox{\cdir/model15T2.eps}
\figcaption[]
{Simulation of model 3.
The age is 180 years.
In this model, $\dot{M}_f=2.5\times10^{-7}$ \smyr{},
$v_{fo}=1000$ \vkm{}, and $\theta_f=10$\degree{}.
The captions are the same as Figure
\ref{fig:model2}, except that 
the dashed line in (\tlabel{a}) is the shell structure derived from
the ballistic bow shock model given by \citet{Ostriker2001}.
\label{fig:model15}
}
\end{figure}
\clearp

\begin{figure} [!hbp]
\centering
\epsscale{0.9}   
\epsfbox{\cdir/model15T2_vrz.eps}
\figcaption[]
{Transverse velocity ($v_R$) and longitudinal velocity ($v_z$) of the shell 
material for model 3 at 180 years. As in Fig. \ref{fig:model2_vrz},
the gray-scale indicates the mass column density on a linear stretch.
The solid lines indicate the $R$ and $z$ components of the mean shell
velocity, while the dashed lines indicate the $R$ and $z$ components 
of the immediate postshock velocity of the AGB wind
derived from the ballistic bow shock model.
\label{fig:model15_vrz}
}
\end{figure}
\clearp

\begin{figure} [!hbp]
\centering
\epsscale{0.6}   
\epsfbox{\cdir/model2P.eps}
\figcaption[]
{Simulation of model 4.
The age is 166 years.
In this model, the fast wind has the same parameters as that in model 1 
but has a time-variation with $A=0.5$ and $P=22$ years 
(see Equation \ref{eq:ft}).
The captions are the same as Figure
\ref{fig:model2}, except that
the dashed lines in (\tlabel{a}) are derived 
from Equation \ref{eq:internal}.
\label{fig:model2P}
}
\end{figure}
\clearp

\begin{figure} [!hbp]
\centering
\epsscale{0.9}   
\epsfbox{\cdir/crl618.eps}
\figcaption[]
{HST images of CRL 618 in [OI] and [SII] emission (logarithmic stretch). 
The cross indicates the approximate position of the central source.
The gray-scale wedges along the right side indicate
the values for the gray-scale images. The intensity of the image
has a unit of erg s$^{-1}$ cm$^{-2}$ arcsec$^2$.
\label{fig:crl618}
}
\end{figure}
\clearp

\begin{figure} [!hbp]
\centering
\epsscale{1}   
\epsfbox{\cdir/model2_emis.eps}
\figcaption[]
{Integrated optical emission and line ratio of
[SII]$\lambda6716$\AA/$\lambda6730$\AA{} for model 1 at an inclination of
$i$=30\degree{}. The optical emission is in 
logarithmic scale with an angular resolution similar 
to that of HST images.
The gray-scale wedges along the right side indicate
the values for the gray-scale images.
\label{fig:model2_emis}
}
\end{figure}
\clearp

\begin{figure} [!hbp]
\centering
\epsscale{1}   
\epsfbox{\cdir/model11_emis.eps}
\figcaption[]
{Same as that in Figure \ref{fig:model2_emis} except for model 2.
\label{fig:model11_emis}
}
\end{figure}
\clearp

\begin{figure} [!hbp]
\centering
\epsscale{1}   
\epsfbox{\cdir/model15T2_emis.eps}
\figcaption[]
{
Same as that in Figure \ref{fig:model2_emis} except for model 3.
\label{fig:model15T2_emis}
}
\end{figure}
\clearp

\begin{figure} [!hbp]
\centering
\epsscale{1}   
\epsfbox{\cdir/model15T2_xray.eps}
\figcaption[]
{Integrated X-ray emission and spectrum for model 3
in the 0.2-1.5 keV energy range at an angular
resolution of 0.5" and an energy resolution of 50 eV.  
The gray-scale wedge along the right side indicates
the values for the gray-scale image.
In the spectrum, there are strong emission lines from oxygen
at about 570 and 653 eV, and from iron at about 730 and 826 eV.
\label{fig:model15T2_xray}
}
\end{figure}
\clearp

\begin{figure} [!hbp]
\centering
\epsscale{1}   
\epsfbox{\cdir/model2P_emis.eps}
\figcaption[]
{
Same as that in Figure \ref{fig:model2_emis} except for model 4.
\label{fig:model2P_emis}
}
\end{figure}
\clearp

\begin{figure} [!hbp]
\centering
\epsscale{1}   
\epsfbox{\cdir/model2_OI.eps}
\figcaption[]
{Position-velocity 
diagrams of the [OI] emission for model 1 at an inclination of
$i$=30\degree{}.
PV diagram on the left has an angular resolution of 0.05" and a velocity
resolution of 10 \vkm{}.
PV diagram on the right is degraded to an angular 
resolution of 1" and a velocity resolution of 50 \vkm{} for comparing with
current groundbased observations.
\label{fig:model2_OI}
}
\end{figure}
\clearp

\begin{figure} [!hbp]
\centering
\epsscale{1}   
\epsfbox{\cdir/model11_OI.eps}
\figcaption[]
{
Same as that in Figure \ref{fig:model2_OI} except for model 2.
\label{fig:model11_OI}
}
\end{figure}
\clearp

\begin{figure} [!hbp]
\centering
\epsscale{1}   
\epsfbox{\cdir/model15T2_OI.eps}
\figcaption[]
{
Same as that in Figure \ref{fig:model2_OI} except for model 3.
\label{fig:model15T2_OI}
}
\end{figure}
\clearp

\begin{figure} [!hbp]
\centering
\epsscale{1}   
\epsfbox{\cdir/model2P_OI.eps}
\figcaption[]
{
Same as that in Figure \ref{fig:model2_OI} except for model 4.
\label{fig:model2P_OI}
}
\end{figure}
\clearp

\begin{figure} [!hbp]
\centering
\epsscale{0.6}   
\epsfbox{\cdir/ratio.eps}
\figcaption[]
{ Line ratios of the [OI]$\lambda$6300\AA, [NII]$\lambda$6583\AA{}
and [OIII]$\lambda5007$\AA{} emission to the
[SII]$\lambda\lambda$6730,6716\AA{} emission.
Solid lines are for number density=2$\times10^3$ \cm3,
dashed lines for 10$^4$ \cm3, and
dotted-dashed lines for 5$\times10^4$ \cm3. The line ratios are derived
assuming equilibrium optically thin emission (see Appendix).
\label{fig:ratio}
}
\end{figure}
\clearp

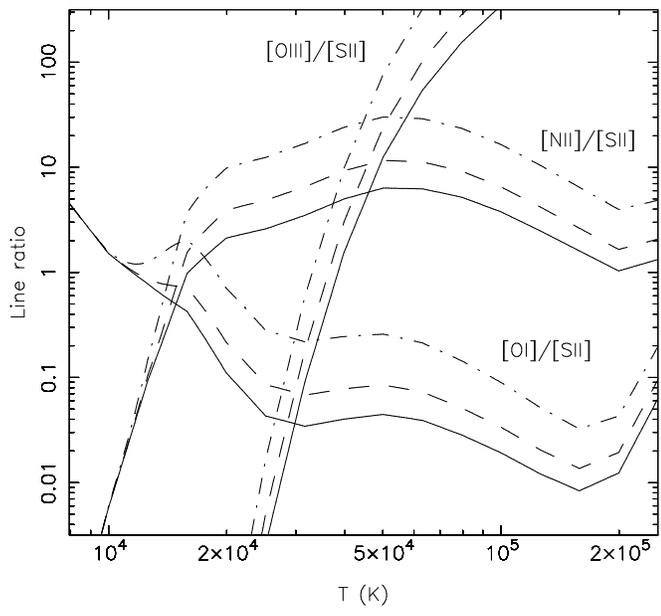
\end{document}